\documentclass[
prl,
twocolumn
]{revtex4}

\usepackage{graphicx}% Include figure files
\usepackage{overpic}% Add letters on images
\usepackage{dcolumn}% Align table columns on decimal point
\usepackage{bm}% bold math
\usepackage[colorlinks=true,linkcolor=red,citecolor=blue]{hyperref}
\usepackage{amsmath}
\newcommand{\tend}[1]{\hbox{\oalign{$\bm{#1}$\crcr\hidewidth$\scriptscriptstyle\bm{\sim}$\hidewidth}}}
\usepackage{ulem}

\begin{document}

\preprint{APS/123-QED}

\title{Thermodynamically consistent phase field model for hydrogen-assisted cracking}

\author{G. F. Bouobda-Moladje}
\thanks{Deceased author}
\affiliation{Université Paris-Saclay, ONERA, CNRS, Laboratoires d'Etudes des Microstructures, Châtillon, France.}

\author{A. Ruffini}
\email{antoine.ruffini@onera.fr}
\affiliation{Université Paris-Saclay, ONERA, CNRS, Laboratoires d'Etudes des Microstructures, Châtillon, France.}

\author{Y. Le Bouar}
\affiliation{Université Paris-Saclay, ONERA, CNRS, Laboratoires d'Etudes des Microstructures, Châtillon, France.}

\author{A. Finel}
\affiliation{Université Paris-Saclay, ONERA, CNRS, Laboratoires d'Etudes des Microstructures, Châtillon, France.}

\date{\today}

\begin{abstract}
We propose a phase field model able to simulate hydrogen-assisted cracking in polycrystalline materials. Within a variational framework, the model simultaneously describes crack propagation and hydrogen segregation on crack surfaces and grain boundaries together with the associated reduction in interfacial energies. In the context of hydrogen-enhanced decohesion (HEDE) mechanisms, we demonstrate the ability of the model to capture the transition from transgranular cracking to hydrogen-assisted intergranular cracking.
\end{abstract}

\maketitle

Hydrogen has been recognized as a promising route to a future with decarbonized energy \cite{peraldobicelli1986,zhang2024}. However, interactions between dissolved hydrogen and microstructural features can lead to catastrophic material degradation and ultimately to unexpected fracture, a phenomenon referred to as hydrogen embrittlement (HE) \cite{nagumo2016,li2020b}. HE of materials is a complex phenomenon and therefore represents a major challenge for hydrogen technologies \cite{campari2023,yu2024hydrogen}. Numerous studies have been dedicated to the fundamental understanding of HE and several mechanisms have been proposed and extensively  reviewed \cite{nagumo2016,li2020b}, including  hydrogen-enhanced decohesion (HEDE), hydrogen-enhanced localized plasticity (HELP), hydrogen-enhanced strain-induced vacancies (HESIV), etc. Due to its multiphysics nature involving several length and time scales, predicting HE using modeling techniques is challenging. A relevant approach to address this type of problems is to develop a model at an intermediate scale between the atomic and macroscopic scales, in which the important mechanisms are taken into account and calibrated to the relevant atomic quantities.

The phase field (PF) method has emerged as the reference approach for modelling the evolution complex systems at the mesoscopic scale and has been successfully applied in the context of solidification \cite{echebarria2004}, phase transformations \cite{onuki2001,finel2010}, electrodeposition \cite{hong2018}, or crack propagation \cite{bourdin2000,karma2001,henry2004}. In the last few years, PF models have been developed for hydrogen assisted cracking by means of a coupling between brittle fracture mechanics and the HEDE mechanism via hydrogen diffusion \cite{martinez-paneda2018,kristensen2020applications,huang2020phase,wu2020phase,chen2022phase,li2022analysis,zhao2024phase,castro2026coupled} Despite their satisfactory validation on case studies, these models need to be improved. First, they do not explicitly reproduce hydrogen segregation, which is the phenomenon responsible for the decrease in  the surface energy of the crack lips and therefore for the HEDE phenomenon. Next, they all relies on the Langmuir-McLean isotherm \cite{langmuir1918,mclean1958,oriani1970diffusion,lejcek2010} to calculate the hydrogen coverage on the surface of the crack based on the local hydrogen concentration. However, this isotherm is a macroscopic equilibrium relationship between homogeneous bulk and surface phases and cannot be justified as a local law, particularly when modeling heterogeneous microstructures involving elastic interactions.

In this paper, we propose a PF model, based on the Kim-Kim-Suzuki (KKS) formalism \cite{kim1999}, to simulate hydrogen-assisted cracking in polycrystalline materials. The KKS formulation is meant to replace the use of a local Langmuir-McLean law for incorporating the local hydrogen concentration. The proposed variational formulation  naturally accounts for the effective crack surface energy degradation induced by hydrogen segregation. The model is first derived within a consistent thermodynamic framework. Then, the effective degradation of the crack surface energy predicted by the model is analytically computed. Finally, the ability of the model to capture the transition from transgranular to intergranular cracking mode assisted by hydrogen \cite{latanision1974,martin2012,ding2021,ding2022} via the HEDE mechanism is investigated.

The generic microstructure considered in this paper consists of a pre-existing crack in a polycrystalline metal that can propagate under mechanical loading. For the sake of simplicity, grain boundaries (GBs) are assumed to remain static, knowing that models accounting for their evolution could be proposed as a natural extension of this work \cite{chen1994,steinbach1996,garcke1999,dimokrati2020}. Hydrogen atoms in solid solution are assumed to occupy available interstitial sites in the bulk and at the 2D microstructural features namely: (i) cracks (with bulk-type surfaces), (ii) intact GBs and (iii) broken GBs (with GB-type surfaces assumed to differ from the bulk-type surfaces).

The state of the system is described by the following fields. First, a damage field $\eta$ related to the material integrity, equal to 1 and 0 respectively for the intact and broken states of the material. Then, the hydrogen occupation probability of the interstitial sites $c$, either in the bulk phase or at the 2D microstructural features. Finally, a set of order parameters $(\phi_1, \phi_2,...,\phi_q)$, each representing a grain of a given crystallographic orientation. Within the grain labelled $i$, the field $\phi_i$ is equal to 1 and and all other fields are equal to 0. As always in continuous phase field modeling, the fields are diffuse so that interfaces or surfaces are represented by a diffuse region.
For simplicity, we will assume in the following that the interstitial sites density in the diffuse regions is equal to that in the bulk.

The total free energy of the system is written as:  
	\begin{align}
		\mathcal{F} =& \int d^{3} \underbar{\textbf{r}} \Big\{ \frac{\gamma^0}{2 C_w} \left[ \frac{w(\eta)}{\xi} + \xi \left|\bm{\nabla} \eta \right|^2 \right]  \nonumber \\
		& + \frac{g(\eta)}{2}  \Big(\tend{\bm{\varepsilon}}-\tend{\bm{\varepsilon}}^{0}(c)\Big):\tend{\tend{\bm{\lambda}}}:\Big(\tend{\bm{\varepsilon}}-\tend{\bm{\varepsilon}}^{0}(c)\Big)   \nonumber  \\ 
				&- p(\eta)\sum_{i<j}\zeta_{ij}\bm{\nabla} \phi_i \cdot  \bm{\nabla} \phi_j  + f_{ch}(c, \eta)   \Big\}.
	\end{align}

The first contribution corresponds to the energy of the surfaces created during crack propagation in a  hydrogen-free material. The surface energy $\gamma^0$ is assumed isotropic. $w(\eta)$ is a cohesion function satisfying $w(1) = 0$, $w(0) = 1$, and $C_w=\int_0^1 \sqrt{w(\eta)} \mathrm d \eta$ is a scaling constant. $\xi$ is the regularization length of the diffuse damage field $\eta$. The second contribution stands for the elastic energy, formulated within the framework of linear elasticity. $\tend{\tend{\bm{\lambda}}}$  denotes the elastic stiffness tensor, $\tend{\bm{\varepsilon}}$ the total strain tensor, $\tend{\bm{\varepsilon}}^{0}(c)$ the eigenstrain tensor associated with the hydrogen field $c$ and  $g(\eta)$ is the material stiffness degradation function with the properties $g(0) = 0$, $g(1) = 1$. The third term accounts for the coupling between the crack and GBs. 
$\zeta_{ij}$ is a parameter that allows us to control the GB fracture energy $\Gamma_{ij}^{GB}$ between grains $i$ and $j$, see Sec. \textcolor{blue}{SII} in the Supplementary Material \textcolor{blue}. The function $p(\eta)$ satisfies $p(0) = 1$, $p(1) = 0$, as well as $p^{'}(1) = 0$, and $p^{''}(1) = 0$ in order to prevent unphysical softening of the GB region, see Sec. \textcolor{blue}{SI} in the Supplementary Material. The last term $f_{ch}$ corresponds to the chemical free energy density, whose formulation constitutes the cornerstone of this work.

In the following, the GB between grains $i$ and $j$ is assumed static and characterized by the profiles $\phi_i = (1+\tanh(2x/\delta_{GB}))/2$ and $\phi_j=1-\phi_i$ where $\delta_{GB}$ is the GB width, and $x$ is the signed distance to the GB. In addition, 
$\tend{\bm{\varepsilon}}^{0}(c)=0$ is assumed for simplicity as the present study mainly focuses on the chemical driving force for hydrogen segregation.

The chemical free energy of the model is written using the so-called KKS approach originally developed in the context of solidification and widely used thereafter \cite{zhou2010,aagesen2017,zhang2023,zeng2024}. The main advantage of this approach is that it enables the interfacial properties to be decoupled from the chemical species concentration fields. Following the KKS approach, we define a bulk phase with site occupancy $c^b$ and a crack surface phase with a site occupancy $c^{ck}$. The chemical energy density $f_{ch}$ and the hydrogen field $c$ are then written as:
\begin{align}
    f_{ch} &= h(\eta)f_{ch}^{b}(c^{b}) + (1-h(\eta))f_{ch}^{ck}(c^{ck}), \nonumber \\
    c &= h(\eta)c^{b} + (1-h(\eta))c^{ck},
\end{align}
where $f_{ch}^{b}$ and $f_{ch}^{ck}$ are the bulk and crack chemical free energy densities, respectively, and $h(\eta)$ is a monotonic function satisfying $h(1) = 1$, $h(0) = 0$. 

Since concentrations $c^{b}$ and $c^{ck}$ are associated with the same chemical species, we must introduce a constraint between these two fields, as there is in fact only one degree of freedom associated with this chemical species. This constraint is the equality of the chemical potentials associated with the free energies of each of the two phases. Here, since hydrogen concentration enters the total free energy only through its chemical component, one has:
\begin{equation}
    \frac{d f_{ch}^b}{d c^{b}} = \frac{d f_{ch}^{ck}}{d c^{ck}},
\end{equation}   
in which $f_{ch}^{\psi}$ are the chemical energy densities, where $\psi$ refers to the bulk ($b$) or crack ($ck$) phase. These quantities are written in the assumption of ideal solid solution:
\begin{equation}
    f_{ch}^{\psi}(c)  = \frac{1}{V_H}\left[E^{\psi}c + k_{\scriptscriptstyle B}T[c\ln c + (1-c)\ln (1-c)]\right],
    \label{eq:sol_regul}
\end{equation}
where $k_{\mathrm{B}}$ is the Boltzmann constant, $T$ is the temperature, and $V_H$ is the atomic volume of a hydrogen atom which is then supposed not to depend on the phase. $E^{\psi}$ is the GB-dependent solution energy per hydrogen atom in the corresponding phase given by
\begin{equation} \label{eq:EbEck}
\begin{aligned}
    E^{b} &= E^{b}_0  - \sum_{i<j}\alpha_{ij}\bm{\nabla} \phi_i.\bm{\nabla} \phi_j, \\
    E^{ck} &= E^{ck}_0  - \sum_{i<j}\beta_{ij}\bm{\nabla} \phi_i.\bm{\nabla} \phi_j,
\end{aligned}
\end{equation}
where $E^{b}_0$ and $E^{ck}_0$ are the solution energy in pure bulk and crack phases, respectively. The hydrogen segregation energy at the crack surface in the single crystal is thus given by $\Delta E^{ck} = E^{ck}_0 - E^{b}_0$. The parameters $\alpha_{ij}$ and $\beta_{ij}$ allow to control respectively the hydrogen segregation energies in the intact GB, $\Delta E^{GB}$, and in the broken GB, $\Delta E^{ck|GB}$ (see Sec. \textcolor{blue}{SIII} in the Supplementary Material).  

The system evolution is assumed to be quasi-static and the governing equilibrium equations are given by:
\begin{equation} \label{eq:kinetics}
\begin{aligned}
    \bm{\nabla}. \tend{\bm{\sigma}} = 0, \\
    \bm{\Delta}\left(\frac{\delta \mathcal{F}}{\delta c}\right) = 0, \\
    \frac{\delta  \mathcal{F}}{\delta \eta} = 0,
\end{aligned}
\end{equation}
where $\tend{\bm{\sigma}}$ is the stress field tensor related to the strain field tensor by
\begin{equation}
    \tend{\bm{\sigma}} = g\left( \eta \right) \tend{\tend{\bm{\lambda}}}:(\tend{\bm{\varepsilon}}-\tend{\bm{\varepsilon}}^{0}(c)).
\end{equation}
Fourier-spectral methods are used for the numerical integration of the equilibrium equations of the hydrogen and damage fields (see \cite{schneider2021review} for a recent review). Mechanical equilibrium is obtained using a fast, robust discrete FFT solver that was recently proposed in Ref.~\cite{finel2025tetrahedron}. 

The cohesion and degradation functions correspond to that of the Karma-Kessler-Levine (KKL) PF model of fracture \cite{karma2001,spatschek2011,mesgarnejad2019}, $w(\eta) = 1 - 4\eta^{3} + 3\eta^{4}$ (giving $C_w = 0.7166$) and $g(\eta) = 4\eta^{3} - 3\eta^{4}$. The interpolation function $p(\eta)$ is given at the lowest order by $p(\eta) = 1 -10\eta^{3} + 15\eta^{4} - 6\eta^{5}$, and the function $h(\eta)$ is simply given by $h(\eta) = 3\eta^{2} - 2\eta^{3}$. 

First, it is useful to analyze the behavior of the model for a planar fracture in the presence of hydrogen. In this one-dimensional geometry, the elastic energy is fully relaxed and, using the regular solutions (Eq.~(\ref{eq:sol_regul})), one obtains the Langmuir–McLean isotherm,
\begin{equation}
\frac{c^{ck}}{1-c^{ck}}=\frac{c^b}{1-c^b} \exp(-\frac{\Delta E^{ck}}{kT}).  
\end{equation}
This law is not assumed to hold at all times along the cracks, but rather emerges here naturally because, under the present condition, the bulk and crack phases are homogeneous.

Next, the variation in surface energy, $\Delta \gamma_{ch}$, induced by segregation can be calculated analytically (see Sec. \textcolor{blue}{SIV} in the Supplementary Material for details of the derivation),
\begin{equation}
    \Delta \gamma_{ch} = C_w^h \,  \xi \, \left( f_{ch}^{ck,eq}-f_{ch}^{b,eq} \right),
    \label{eq:delta_gamma}
\end{equation}
where $C_w^h = \int_0^1 \left( 1 - h(\eta) \right) / \sqrt{w(\eta)} \mathrm d \eta$ is an integration constant of the order of unity. $f_{ch}^{ck,eq}$ and $f_{ch}^{b,eq}$ are the chemical energy densities of the crack and bulk phases, respectively. In the limit of low hydrogen concentrations, this expression becomes
\begin{equation}
	\Delta \gamma_{ch} =  C_w^h \  f_{ch}^{b}  \ {\xi} \ \left( \exp \left[ -\frac{\Delta E^{ck}}{k_B T} \right] -1 \right). 
    \label{eq:delta_gamma2}
\end{equation}
This relation determines the value of the parameter $\xi$, which is related to the surface crack thickness, through a given surface energy defined as a function of the hydrogen coverage curve \cite{alvaro2015,katzarov2017}.

We now perform a numerical validation of the model by measuring the critical stress $\sigma_c$ for crack growth in mode I under uniform applied stress. The simulations are performed in 2D without and with the HEDE mechanism using the parameters listed in Table \ref{data_Al}.
\begin{table}[!h]
  \begin{center}	
	\caption{\label{data_Al} PF parameters for validation in the case of an isotropic model material with Young modulus $E$ = 80 GPa and Poisson ratio $\nu$ = 1/3. Hydrogen bulk properties correspond to that in Aluminum.}
    \begin{tabular}{ll}
			\hline \\ 
			$d$& 0.4 $\mu$m \\
			$T$   & 300 K \\
			$\xi$, $\gamma^{0}$ & 4d, 20 J/m$^2$   \\
			$V^{\psi}$ & 16 \AA$^{3}$ \cite{kirchheim2014} \\ 
			%$G_0^b$, $\Delta G_{0}$ & 0.68 eV, -0.6 eV \\
			$\Delta E^{ck}$ & -0.6 eV \cite{huang2017,he2019,drexler2021} \\
			$c^{b, eq}=\exp \left( -E_0^b/(k_{\scriptscriptstyle B} T) \right) $  &  $1.5\times10^{-12}$ \cite{ambat1996} \\  
			\hline 	   
	\end{tabular}
 \end{center}
\end{table}

\begin{figure}[b]
	\includegraphics[width=0.9\linewidth]{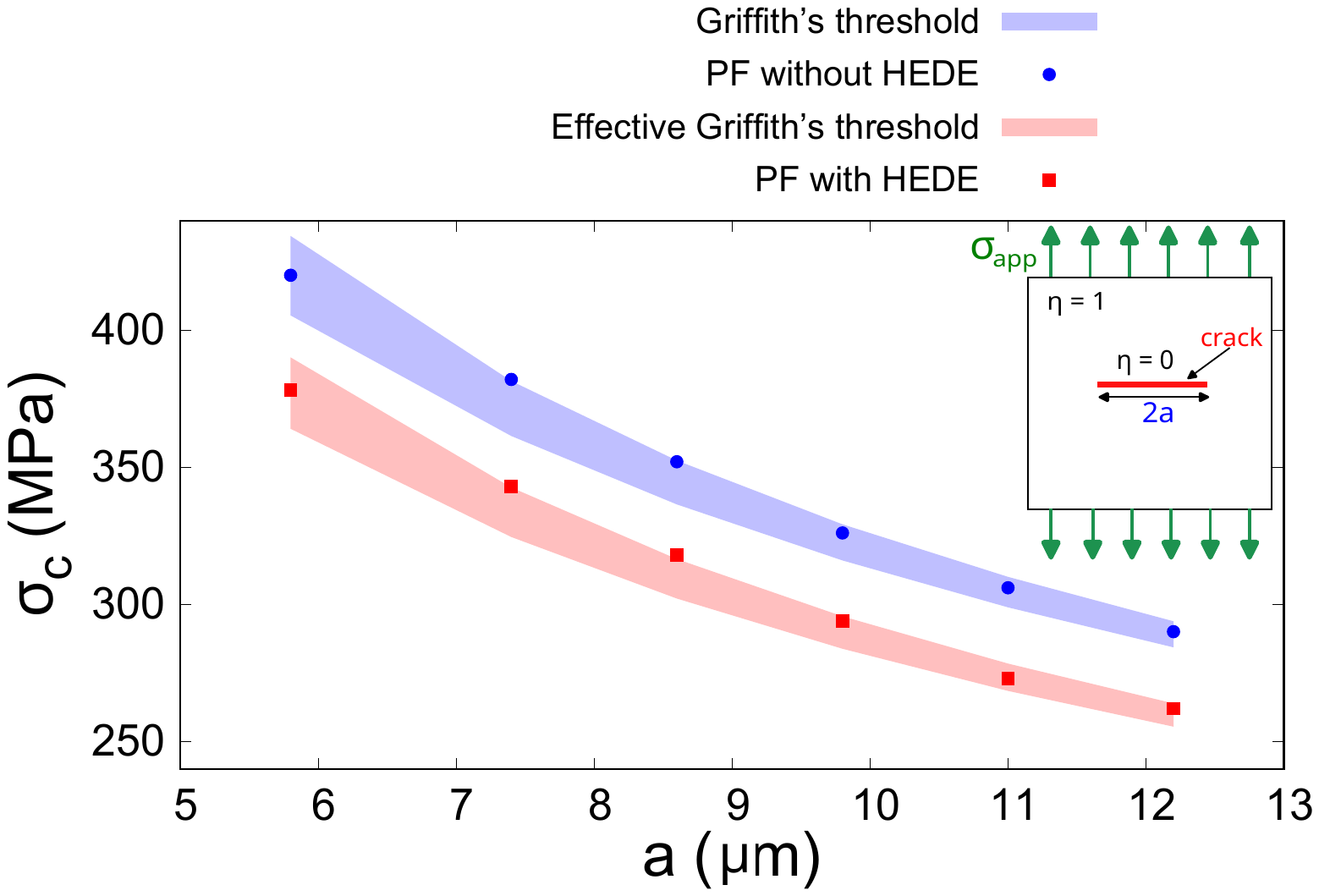}
	\caption{\label{citical stress} Critical stress $\sigma_c$ for crack propagation in mode I as a function of the initial half-length $a$ of the crack in a single crystal. 
	The reduction of $\sigma_c$ with the crack length as well as with the hydrogen content is in good agreement with the Griffith's criterion (colored regions).}
\end{figure}

The PF results without the HEDE mechanism (blue plain circles in Fig. \ref{citical stress}), show a good agreement with the Griffith theoretical criterion $\displaystyle \sigma_{c}^{th} = \sqrt{\frac{2E\gamma^{0}}{\pi (a \pm \Delta a)}}$ \cite{1921,sirtori1992}, with $E$ and $a$ the Young's modulus and the initial half-length of the crack, respectively. Because the crack field is discretized for the numerical implementation, the crack length entering in the Griffith's criterion inherits an uncertainty $\Delta a$ which is on the order of the grid spacing.       
By turning on the HEDE mechanism within the PF simulations, we observe a decrease of $\sigma_c$ (red plain squares in Fig.~\ref{citical stress}) due to a hydrogen-induced surface energy variation $\Delta \gamma_{ch}$ involved in the energy balance during crack growth. Indeed, in the presence of hydrogen, the results obtained within the phase-field model still follow the Griffith criterion, but with an effective surface energy reduced by about 4 J·m$^{-2}$. This result is consistent with the value $\displaystyle \Delta \gamma_{ch} = -4.29$ J·m$^{-2}$ obtained using equations (\ref{eq:delta_gamma}) or (\ref{eq:delta_gamma2}).

Next, the model is applied in the context of hydrogen assisted crack propagation in polycrystalline materials. H-induced intergranular cracking occurs due to the susceptibility of GBs to HE mechanisms. While it is experimentally difficult to quantify the contribution of each mechanisms, due to their synergistic nature \cite{djukic2019,liang2021}, H-assisted intergranular cracking is commonly interpreted as the loss of GB toughness due to the presence of hydrogen \cite{mcmahonjr.2004,robertson2015,quan2025new}. Atomistic simulations have been extensively used to assess the reduction in cohesive properties induced by H segregation \cite{alvaro2015,huang2017,he2019,li2020a}. Notable measurement is the reduction of the GB strength by 50\% in the slow fracture limit in Ni, obtained by combining atomistic calculations and thermodynamic theory \cite{huang2017}. It has been demonstrated theoretically using linear elastic fracture mechanics that the crack deflection at weak region like GB depends on the GB/bulk fracture energy ratio and the angle of deflection \cite{ming-yuan1989}. PF simulations have been also used to predict the crack grow path across GB \cite{zeng2017,nguyen2017a,henry2019,chen2019}, pointing out the role of the applied strain \cite{chen2019} which is not considered in the theoretical prediction. Here, we investigate quantitatively the crack deflection at GB, related to H-induced decrease of the GB/bulk fracture energy ratio (HEDE mechanism), in a polycrystalline metal. For the sake of simplicity, we assume isotropic bulk fracture energy $G_c^0 = 2\gamma^0$ inside the grains, uniform distribution of the GB fracture energy $\Gamma^{GB}$, and of the H segregation energies $\Delta E^{GB}$ and $\Delta E^{ck|GB}$ in intact and broken GBs, respectively. The simulations are performed in 2D using the parameters listed in Table \ref{data_Ni}.
\begin{table}[!h]
	\begin{center}	
		\caption{\label{data_Ni} PF parameters for model application in a polycrystalline material. The material properties correspond to that of Nickel with Young modulus $E$ = 209 GPa and Poisson ratio $\nu$ = 0.30 (Reuss averaging of the cubic constants: C$_{11}$ = 245 GPa, C$_{12}$ = 140 GPa, C$_{44}$ = 125 GPa).}
		\begin{tabular}{ll}
			\hline \\ 
			$d$& 0.352 nm \\
			T   & 700 K \\
			$\xi$, $\delta_{\mathrm{\scriptscriptstyle GB}}$,  $\gamma^{0}$ & 4$d$, 2$d$, 20 J/m$^2$   \\
			$\displaystyle V^{\psi}$ & 2.4 \AA$^{3}$ \cite{griessen1985}  \\
			$\Delta E^{ck}$, $\Delta E^{GB}, \Delta E^{ck|GB}$ & -0.5 eV, -0.25 eV, -0.9 eV   \\
		    $c^{b, eq}=\exp \left( -E_0^b/(k_{\scriptscriptstyle B} T) \right) $  &  $1.5\times10^{-4}$ \cite{wipf2001} \\  
			\hline 	   
		\end{tabular}
	\end{center}
\end{table}

The set of functions $\left\{ w(\eta),g(\eta),p(\eta),h(\eta) \right\}$ defined above are used. The loading conditions correspond to a mode I for the pre-existing crack. The hydrogen segregation energies parameterization is based on atomistic calculations performed in Nickel \cite{huang2017,he2019,drexler2021}: $\Delta E^{ck} = -0.5$ eV,  $\Delta E^{GB} = -0.25$ eV, and $\Delta E^{ck|GB} = -0.9$ eV.

\begin{figure}
\begin{overpic}[trim=90 30 100 0, clip, width=0.48\linewidth]{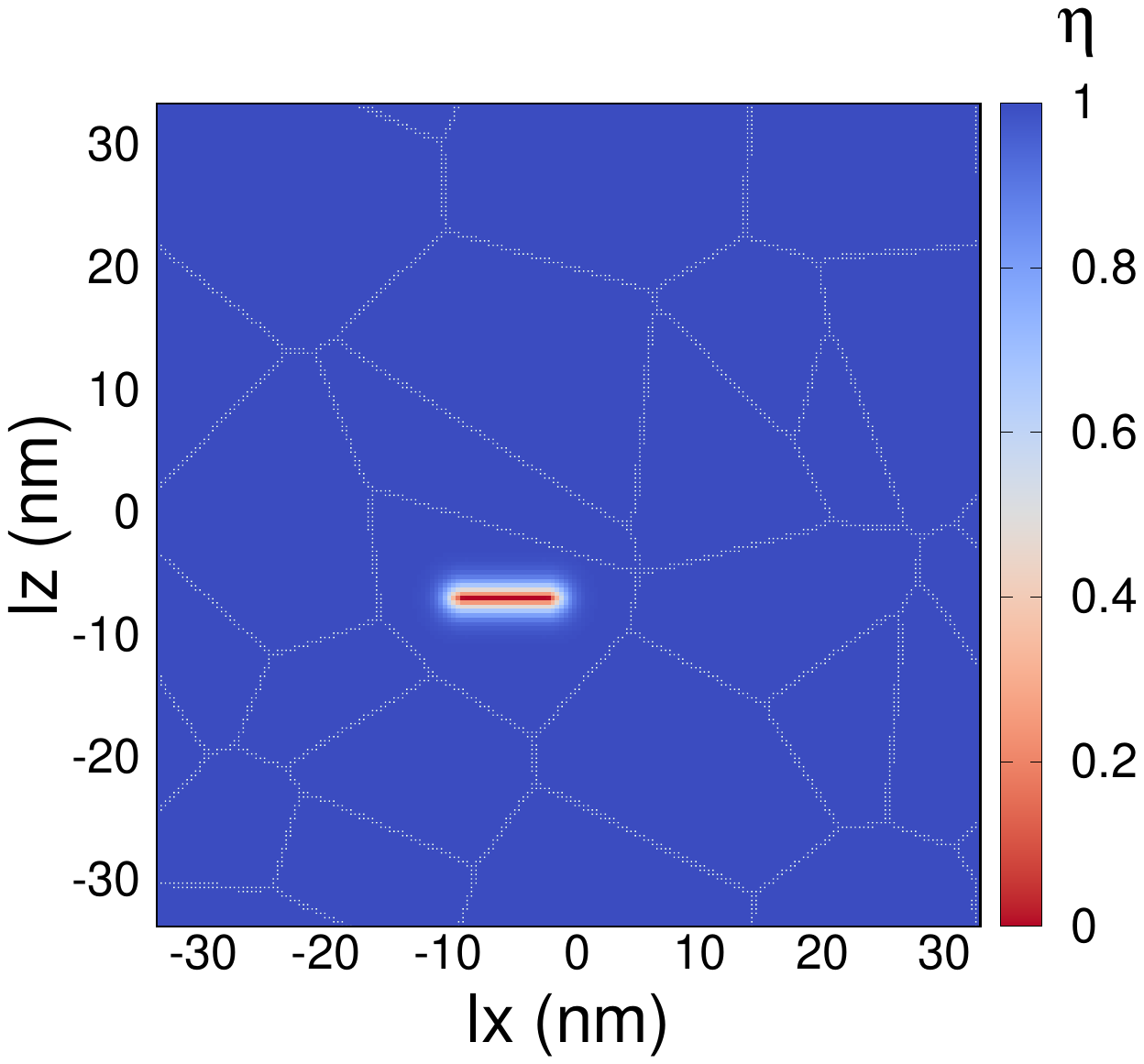}
    \put(5,80){\textbf{a)}}
\end{overpic}
\begin{overpic}[trim=90 30 100 0, clip, width=0.48\linewidth]{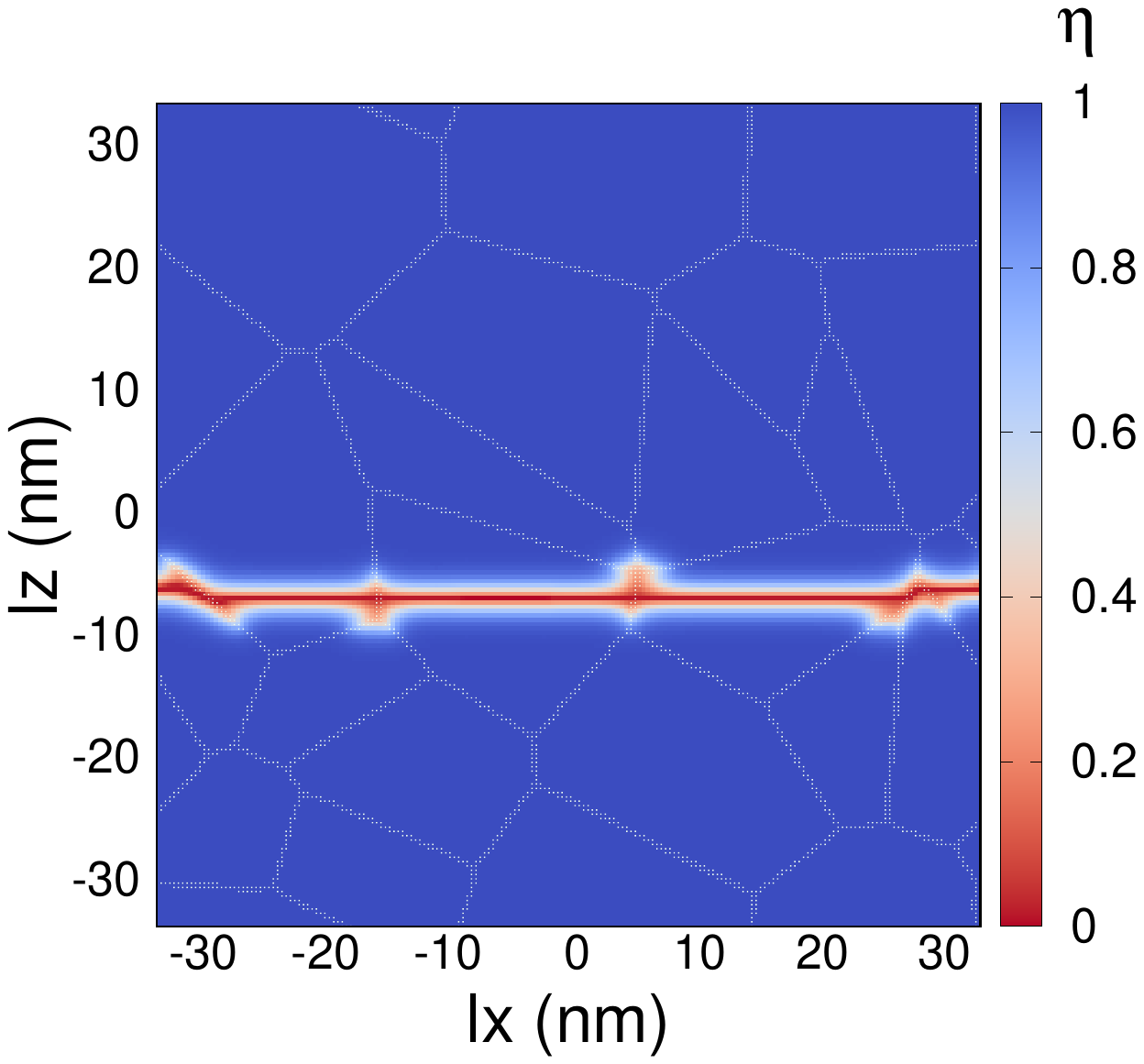}
    \put(5,80){\textbf{b)}}
\end{overpic}
\begin{overpic}[trim=90 30 100 0, clip, width=0.48\linewidth]{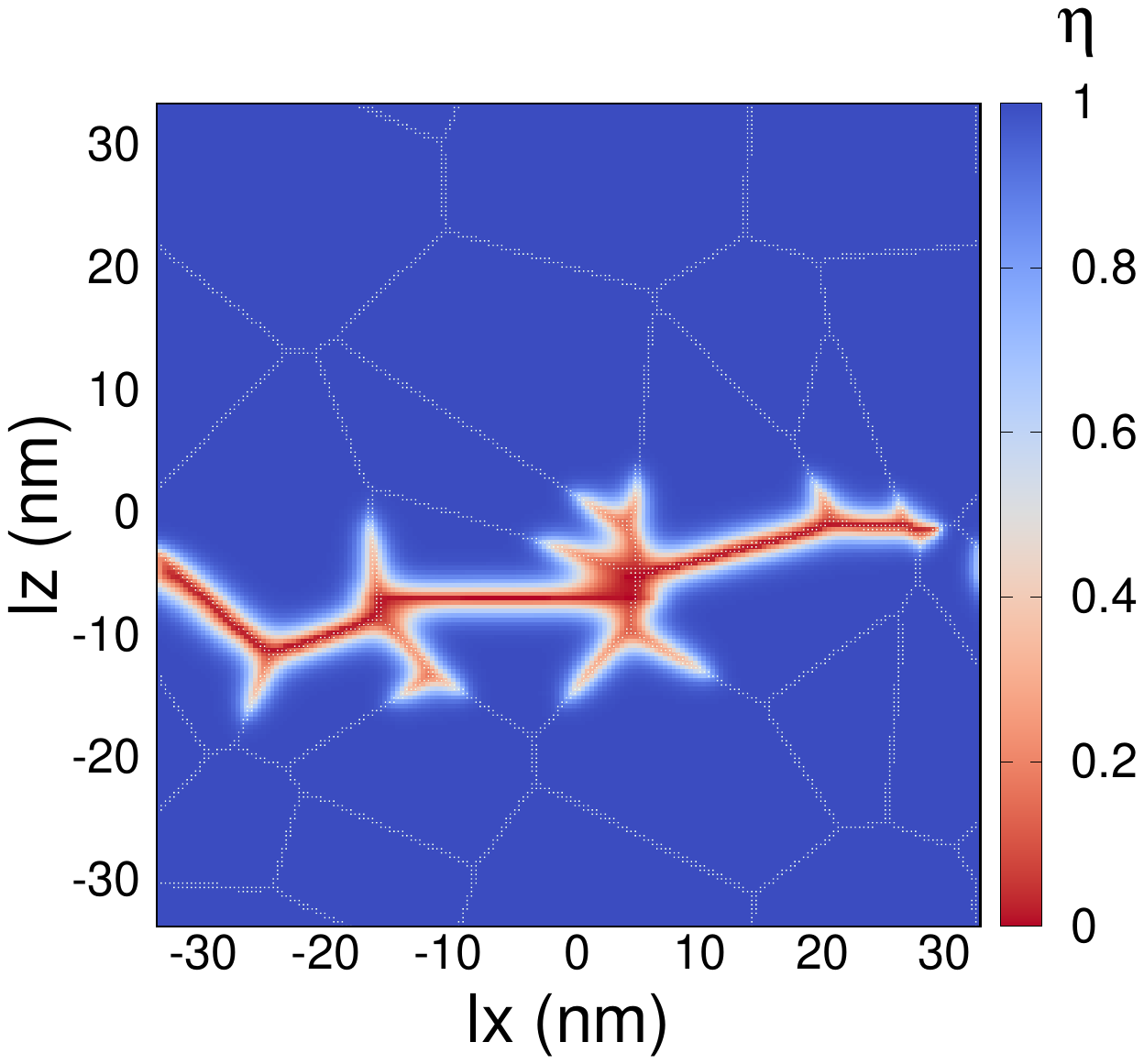}
    \put(5,80){\textbf{c)}}
\end{overpic}
\begin{overpic}[trim=90 30 100 0, clip, width=0.48\linewidth]{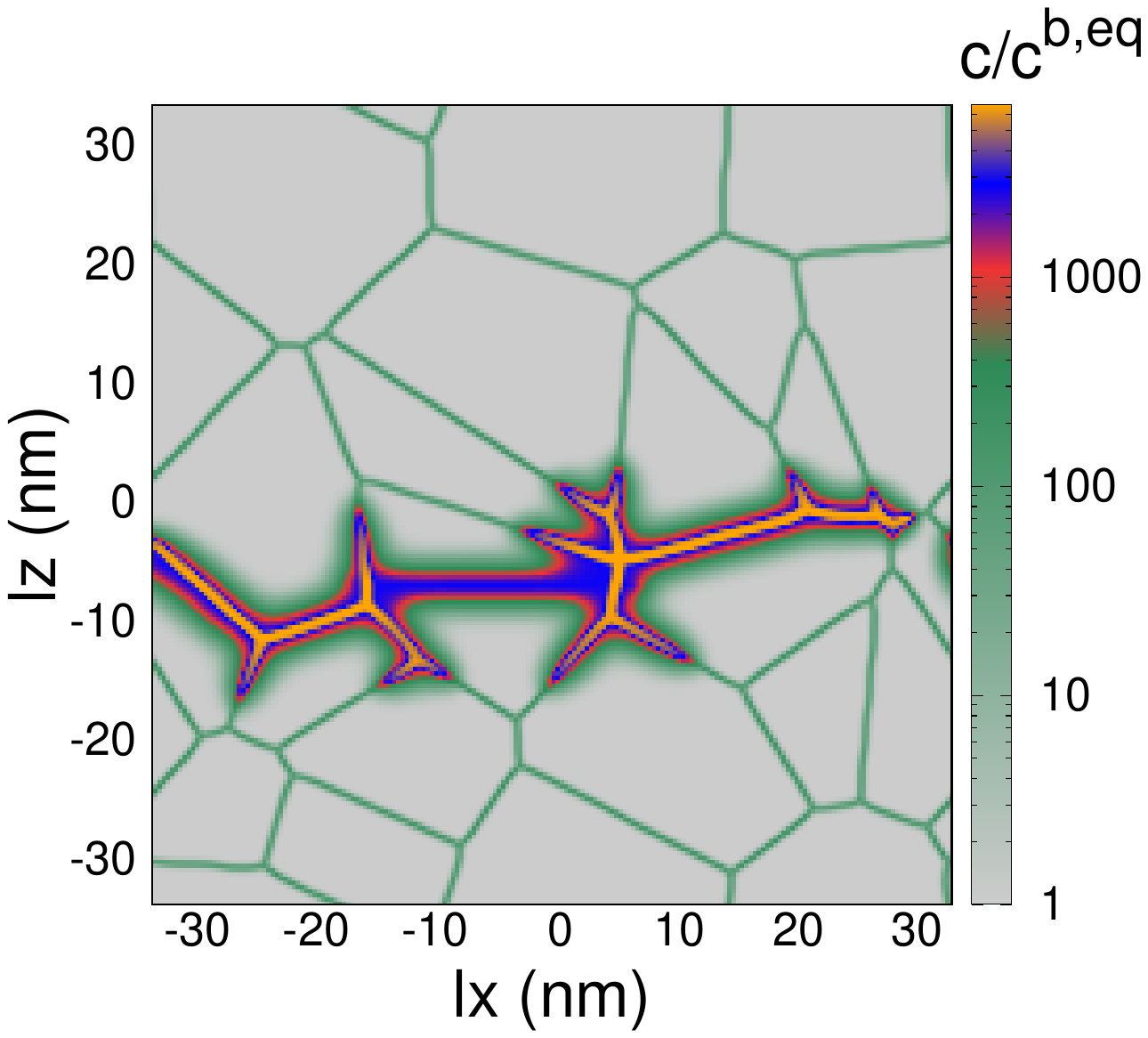}
    \put(5,80){\textbf{d)}}
\end{overpic}
	\caption{\label{crack_path_poly} (a) initial crack configuration in the polycrystalline system. Crack growth path showing the transition from transgranular (b) to intergranular (c) cracking mode respectively without and with the HEDE mechanism. (d) Corresponding hydrogen concentration field normalized by the bulk equilibrium concentration.}
\end{figure}

The reference case corresponds to a hydrogen-free material containing a pre-exiting crack. The GB/bulk fracture energy ratio is set to $\Gamma^{GB}/G_c^0 = 0.5$. The crack propagation path is only transgranular in this case (high-toughness GBs), see Fig. \ref{crack_path_poly}.b. The comparison of the crack path without and with hydrogen in the system clearly shows the transition from transgranular to intergranular cracking mode, see Fig. \ref{crack_path_poly}.b and c. The PF measurement of the GB/bulk fracture energy ratio indicates its decrease with the hydrogen presence; i.e. $(\Gamma^{GB}/G_c^{0})_{\mathrm{eff}} = 0.07 $. More precisely, both the bulk and GB fracture energies decrease with the hydrogen coverage at the bulk-type surfaces of cracks and at the GB-type surfaces of broken GBs (see Sec. \textcolor{red}{SV} in the Supplementary Material). Since the H segregation energy at GB-type surface is higher than that at the bulk-type surface, the decrease of the GB fracture energy is more important than that of the bulk. This results in a drastic reduction of the GB/bulk fracture resistance ratio with hydrogen content, and hence to the observation of the intergranular cracking mode. The map of the hydrogen field, see Fig. \ref{crack_path_poly}.d, reveals the contrast between the H segregation energies at GB, at the crack surface within the grain, and at the broken GB surface.

To summarize, we have developed a new PF model coupling the linear elastic fracture mechanics to the HEDE mechanism within a KKS formalism. The present model reproduces hydrogen segregation at the crack surface in a variationally consistent thermodynamical framework without recourse to the Langmuir–McLean isotherm relationship, which pertains exclusively to thermodynamic equilibrium between macroscopic homogeneous phases and, therefore, cannot be used in heterogeneous situations. This law is yet shown to remain verified in the limit of the static stress-free one dimensional case because this specific situation leads to an equilibrium between homogeneous bulk and crack phases. The decrease in surface energy as a function of hydrogen content is thus reproduced rigorously and generically, which proves particularly well-suited for modeling the HEDE mechanism at the microstructural scale. An application to H-assisted cracking in a polycrystalline material notably shows its capability to capture the transition from transgranular to intergranular cracking mode induced by the HEDE mechanism when using realistic input data for nickel. In this work, surface energy anisotropy, which can affect crack propagation (see for example \cite{perez2000,takei2013,liu2022}), has been neglected. Anisotropy effects may be taken into account through higher-order term in the functional energy (see for example \cite{abinandanan2001,li2015,li2019}), but this is left for future work. Upcoming efforts will be devoted to the incorporation of the others HE mechanisms in the present model and to the investigation of their synergistic effects on the crack propagation path. For the HELP mechanism, a first step would consist in the introduction of individual dislocations in the system using PF model of dislocations \cite{rodney2000phase,wang2001,rodney2003,geslin2015,ruffini2017}. It will therefore require to develop a consistent thermo-elastic and kinetic coupling between dislocation, cracks and hydrogen fields.

%\begin{acknowledgments}
%
%\end{acknowledgments}

\nocite{*}
\bibliography{biblio_HE_fracture_PF}% Produces the bibliography via BibTeX.

\end{document}

\maketitle
	
\renewcommand{\thetable}{S\arabic{table}}	
\renewcommand{\theequation}{S\arabic{equation}}
\renewcommand{\thefigure}{S\arabic{figure}} 	

\section*{SI: Influence of the coupling function $p(\eta)$ on the material integrity}
	
\begin{figure}[h]
	\centering
	\begin{subfigure}{0.5\textwidth}
		\includegraphics[trim = 100 10 100 10, width=\textwidth]{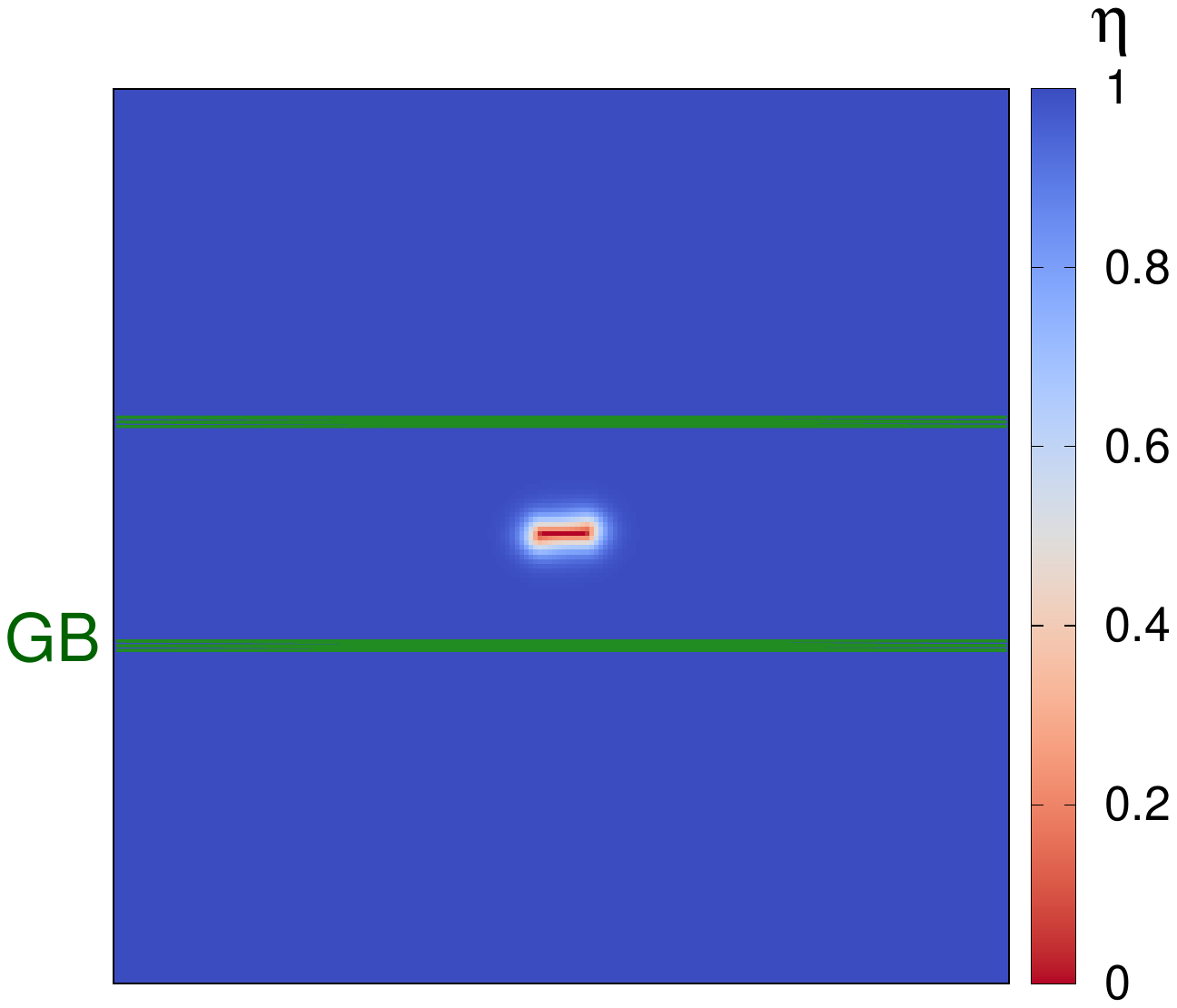} %  gauche bas droite haut
		\caption{Initial state}
	\end{subfigure}
	\begin{subfigure}{0.5\textwidth}
		\includegraphics[trim = 100 10 100 10,width=\textwidth]{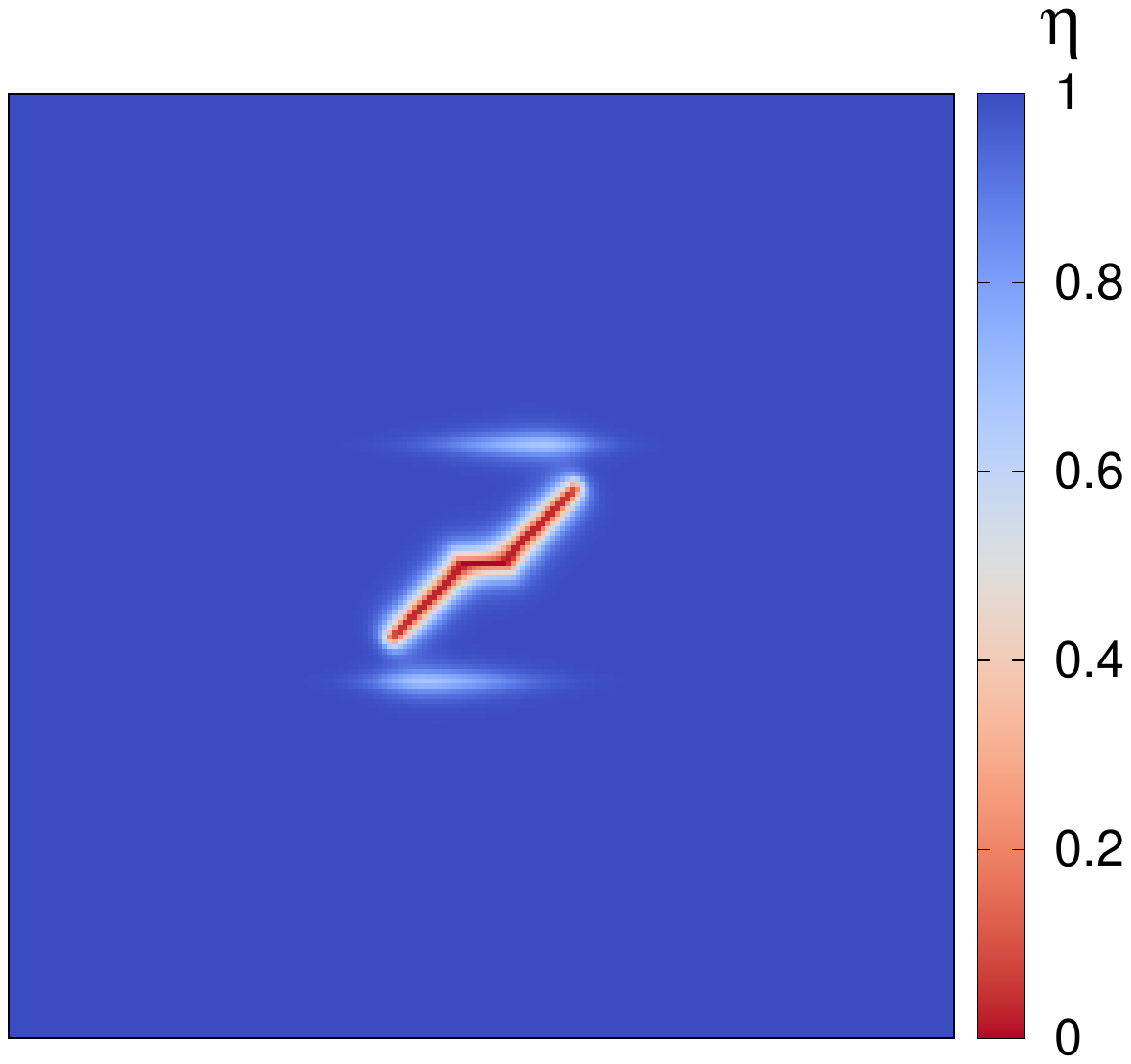}
		\caption{$p(\eta) = 1 - 4\eta^{3} + 3\eta^{4}$, state 1}
	\end{subfigure}%
	\begin{subfigure}{0.5\textwidth} 
		\includegraphics[trim = 100 10 100 10,width=\textwidth]{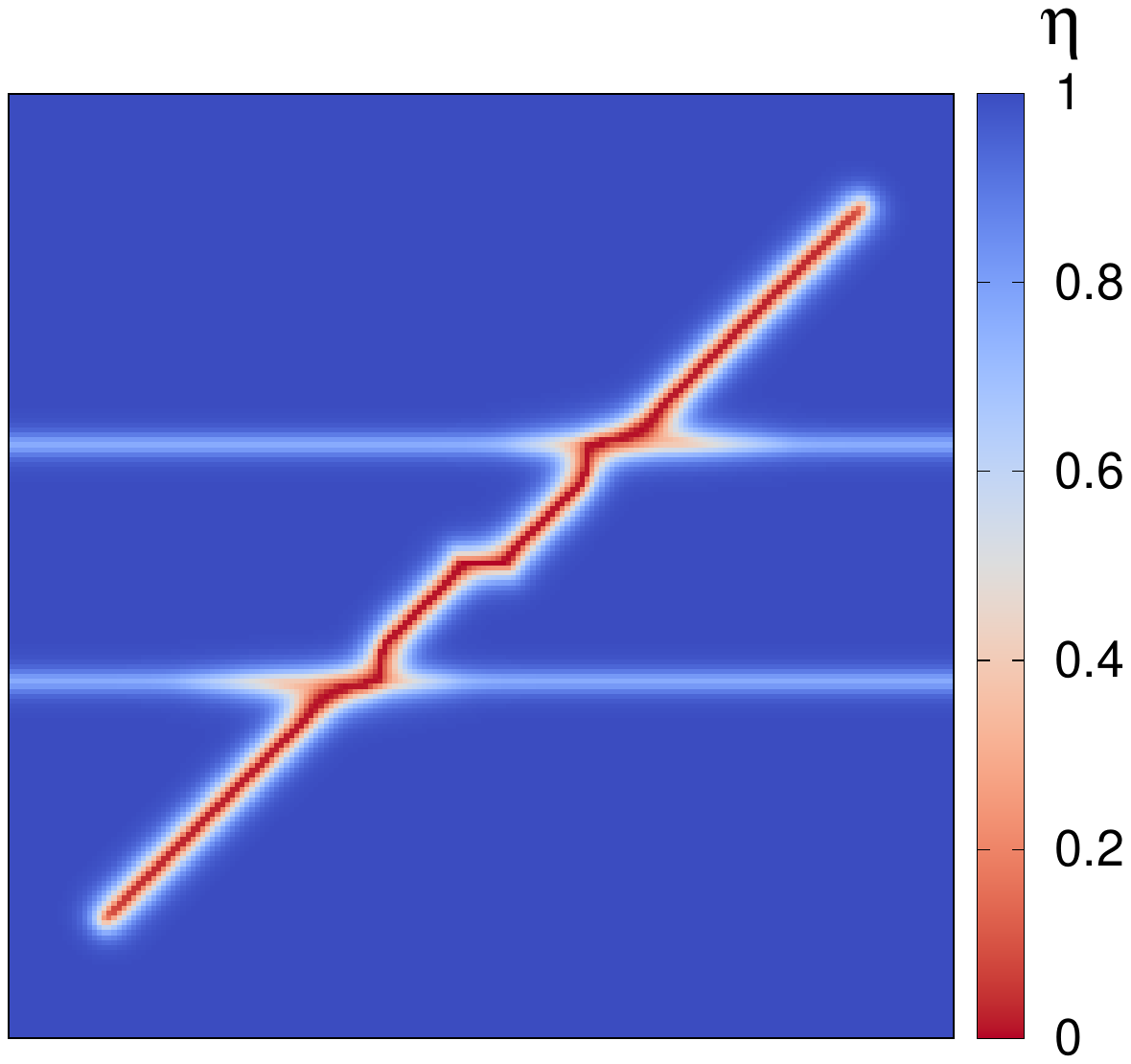}
		\caption{state 2}
	\end{subfigure}
	\begin{subfigure}{0.5\textwidth}
		\includegraphics[trim = 100 10 100 10,width=\textwidth]{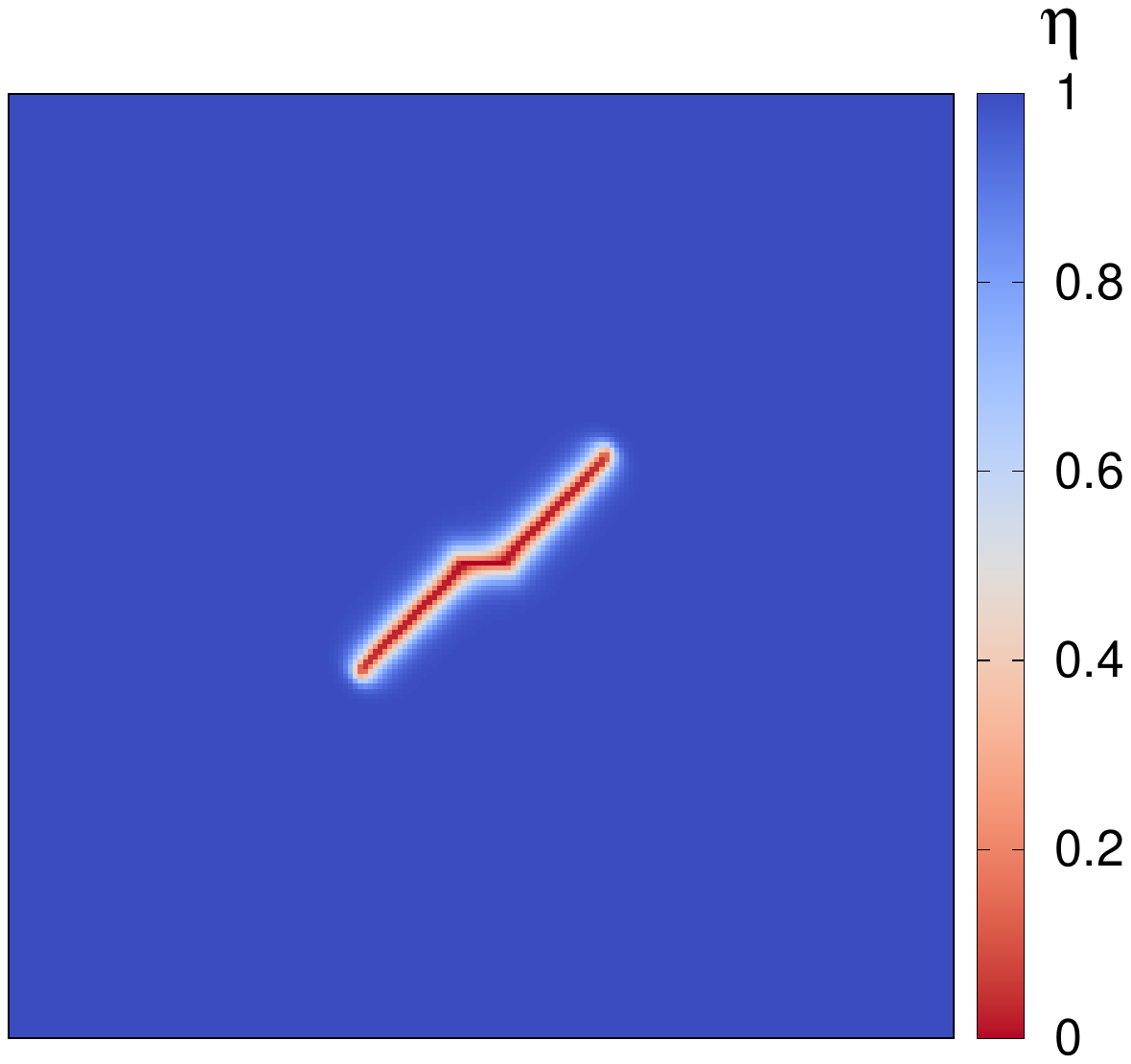}
		\caption{$p(\eta) = 1 -10\eta^{3} + 15\eta^{4} - 6\eta^{5}$, state 1}
	\end{subfigure}%
	\begin{subfigure}{0.5\textwidth} 
		\includegraphics[trim = 100 10 100 10,width=\textwidth]{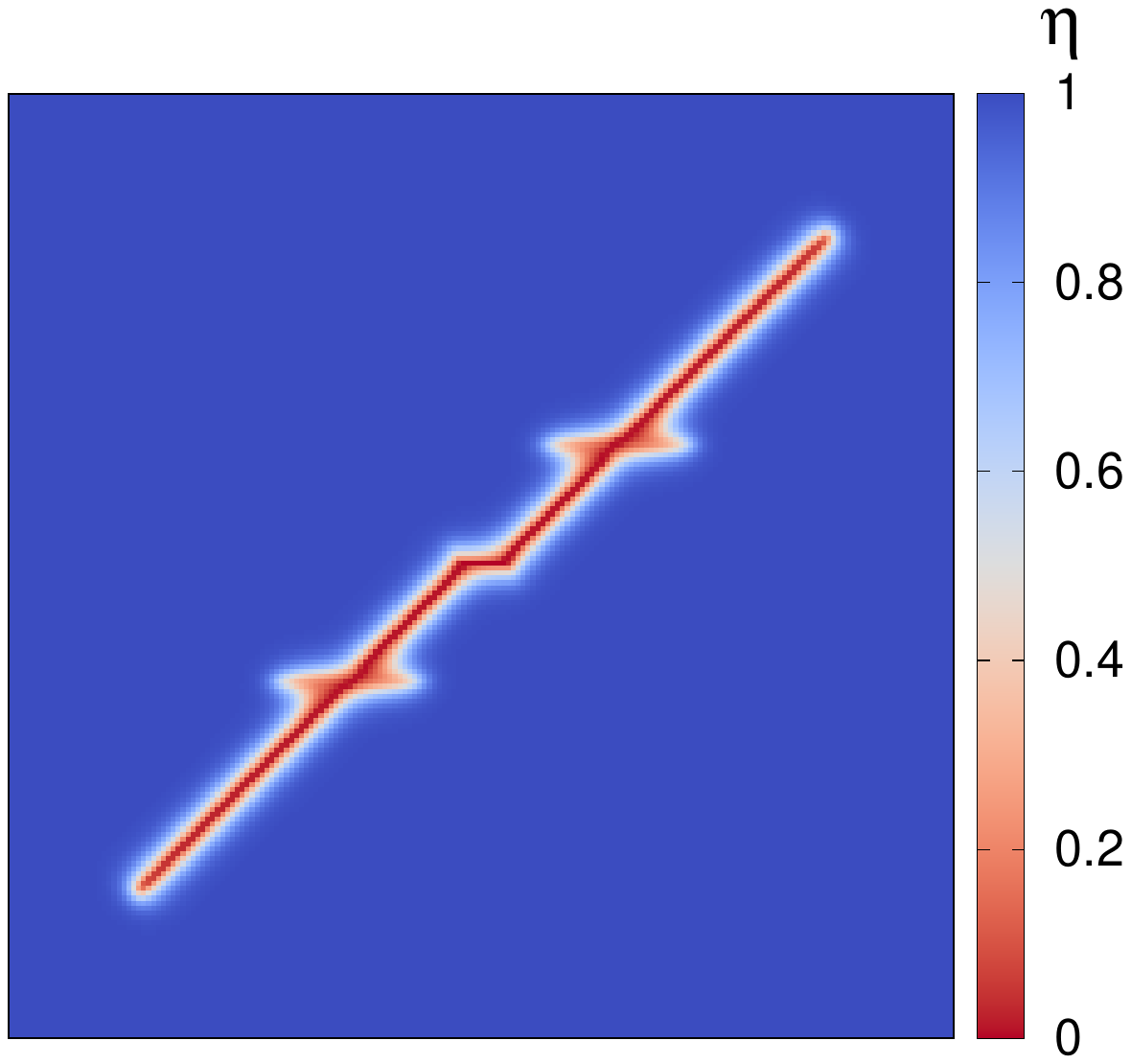}
		\caption{state 2}
	\end{subfigure}
	\caption{Crack propagation path for different choices of the interpolation function $p(\eta)$. The function $p(\eta) = 1 - 4\eta^{3} + 3\eta^{4}$ induces nonphysical softening of the GB region which does not occur when $p(\eta) = 1 -10\eta^{3} + 15\eta^{4} - 6\eta^{5}$.}
	\label{fig_fct_p}
\end{figure}

In the initial state, a crack is introduced in the middle of 2D system in which two horizontal grain boundaries (GBs) are considered (see Fig. \ref{fig_fct_p}.a). A $45^\circ$ loading is applied making the crack propagates across the GBs. When the interpolation function $p(\eta) = 1 - 4\eta^{3} + 3\eta^{4}$ is used, an nonphysical softening of the GB region is observed, even before crack passing across the GB  (see Fig. \ref{fig_fct_p}.b and c). This nonphysical softening is removed by using the interpolation function $p(\eta) = 1 -10\eta^{3} + 15\eta^{4} - 6\eta^{5}$ whose second derivatives vanishes for $\eta =0$ and $\eta=1$ (see Fig. \ref{fig_fct_p}.d and e).

\newpage

\section*{SII: Link between the PF parameter $\zeta_{ij}$ and the GB fracture energy  $\Gamma_{ij}^{GB}$}	

\begin{figure}[ht]
	\centering
	\begin{subfigure}[b]{0.9\textwidth}
		\includegraphics[width=\textwidth]{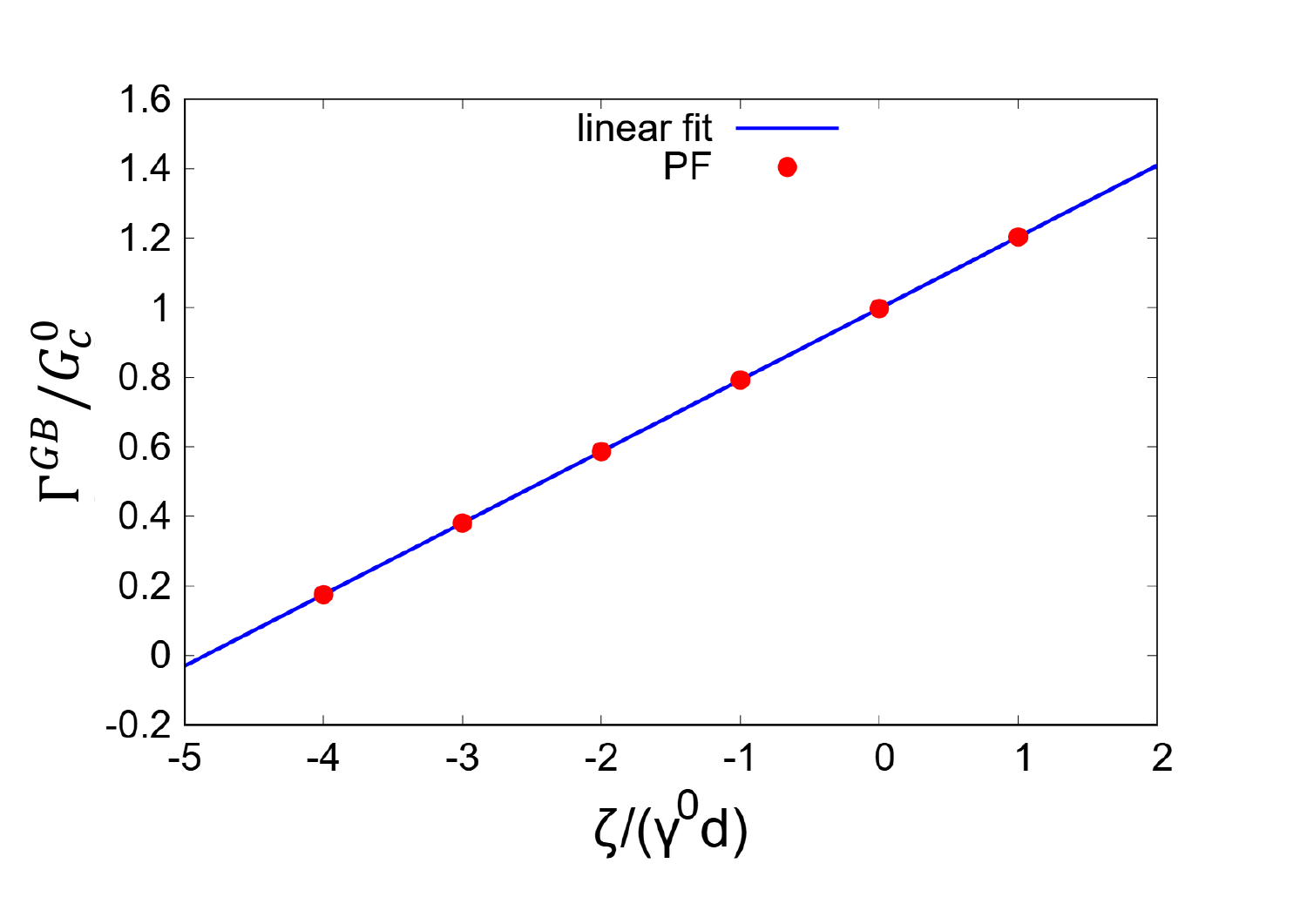}
		\caption*{}
	\end{subfigure}
	\caption{Linear evolution of the GB/bulk fracture energy ratio $\Gamma^{GB}/G_c^0$ with the parameter $\zeta$.}
	\label{nrj_adh}
\end{figure}

By definition, in the absence of hydrogen, the GB fracture energy $\Gamma_{ij}^{GB}$ of the GB separating grains $i$ and $j$ corresponds to the energy difference between the broken GB and the virgin state,
\begin{equation}
	\Gamma^{GB} =  \frac{1}{\mathcal{A}} \int_{V} \left( \frac{G_c^0}{4 c_w}\left[\frac{w(\eta)}{\xi} + \xi \left|\bm{\nabla} \eta \right|^{2}\right] - \zeta \, p(\eta)\, \bm{\nabla} \phi_i \cdot \bm{\nabla} \phi_j   \right) \mathrm{d} V,
\end{equation}
where the subscript $ij$ indicating the type of grain boundary has been dropped for clarity
and where $\mathcal{A}$ is the total area of the broken GB (we recall that the fracture energy is linked to the surface energy by $G_c^0 = 2 \gamma^0$). For a symmetric and planar GB perpendicular to the $x$ axis, one has $\phi_i(x)=\phi(x)=1-\phi_j(x)$, so that
\begin{equation}
	\Gamma^{GB} =  \int_{-\infty}^{+\infty} \left( \frac{G_c^0}{4 c_w}\left[\frac{w(\eta)}{\xi} + \xi \left|\frac{d \eta}{d x} \right|^{2}\right] + \zeta \, p(\eta)\left|\frac{d \phi}{d x}\right|^{2}  \right) \mathrm{d} x.
\end{equation}
When the damage field $\eta$ covering the GB is assumed to be close to the solution $\eta_0$ -- making the first term to be equal to $G_c^0$ by construction of the phase field fracture model -- one has
\begin{equation}
	\Gamma^{GB} \approx G_c^0  + \zeta \int_{-\infty}^{+\infty} p(\eta)\left|\frac{d \phi}{d x}\right|^{2}  \mathrm{d} x.
\end{equation}
By introducing $\tilde x = x/d$ and using the profiles' symmetry, one finally gets
\begin{equation}
	\frac{\Gamma^{GB}}{G_c^0} \approx 1  + \frac{\zeta}{\gamma^0 d} \int_{0}^{+\infty} p(\eta) \left|\frac{d \phi}{d \tilde x}\right|^{2}  \mathrm{d} \tilde x.
\end{equation}
testifying of a linear relationship between $\Gamma^{GB}/G_c^0$ and $\zeta/(\gamma^0 d)$ with a slope provided by the integral. This one is finite and depends on the choice of the function $p(\eta)$, the damage profile $\eta$ and the grain profile $\phi$. In this work, we chose $p(\eta) = 1 -10\eta^{3} + 15\eta^{4} - 6\eta^{5}$, the damage profile $\eta=\eta_0(\xi)$ is that of the KKL phase field fracture model involving $w(\eta) = 1 - 4\eta^{3} + 3\eta^{4}$ and we use $\phi = (1+\tanh(2x/\delta_{GB}))/2$. When considering $\xi = 4d$ and $\delta_{GB} = 2d$, the integral is equal to $0.16$. This slope can be numerically evaluated by a simple one-dimensional simulation of the planar broken GB in which the damage profile $\eta$ differs from $\eta_0$ as assumed previously. In this case, a linear relationship is also found but the slope is closer to $0.20$ (see Fig. \ref{nrj_adh}). In our study, we used this numerical estimation by taking $\zeta/(\gamma^0 d) = -2.5$ to set $\Gamma^{GB}/G_c^0 = 0.5$.

\newpage

\section*{SIII: Link between the PF parameters $\alpha_{ij}$, $\beta_{ij}$ and the hydrogen segregation energies in intact and broken GB}

In this section, we also consider a symmetric and planar GB between grains $i$ and $j$ and the indices are dropped for clarity ($\alpha_{ij} \rightarrow \alpha$ and $\beta_{ij} \rightarrow \beta$). We define an effective hydrogen segregation energy in an intact GB as the average excess solution energy in the GB,
\begin{equation}
	\Delta E^{GB} = \frac{1}{\delta_{GB}}\int \left(E_0^b + \alpha \left(\frac{d \phi}{d x} \right)^{2}\right) -  E_0^b \ \mathrm{d}x,
\end{equation}
where the GB diffuse profile $\phi = (1+\tanh(2x/\delta_{GB}))/2$ only spreads along the $x$ axis over a distance of the order of $\delta_{GB}$.  When considering continuous fields, this quantity can be explicitly calculated:
\begin{align}
	\Delta E^{GB} 
	%&= \frac{1}{V_{GB}}\int \left(E_0^b + \alpha \left( \frac{\partial \phi(x)}{\partial x}\right)^{2}\right)  \mathrm{d} V -  E_0^b \nonumber \\
	%&=\frac{\alpha}{V_{GB}} \int \left( \frac{d \phi(x)}{d x}\right)^{2}  \mathrm{d} V \nonumber \\
	&=\frac{\alpha}{\delta_{GB}} \int \left( \frac{d \phi(x)}{d x}\right)^{2}  \mathrm{d} x, \nonumber \\
	&=\frac{\alpha}{\delta_{GB}} \int_{0}^{1} \frac{d \phi(x)}{d x}  \mathrm{d} \phi, \nonumber \\
	&=\frac{\alpha}{\delta_{GB}} \int_{0}^{1} \frac{4 \phi \left(1-\phi\right)}{\delta_{GB}}  \mathrm{d} \phi, \nonumber \\
	&=\frac{2\alpha}{3\delta_{GB}^2}.
\end{align}

\noindent Similarly, the hydrogen segregation energy in a broken GB is defined by
\begin{align}
	\Delta E^{ck|GB} &= \Delta E^{ck} + \frac{1}{\delta_{GB}}\int  \beta \left( \frac{\partial \phi(x)}{\partial x}\right)^{2}   \ dx, \nonumber \\
	&=\Delta E^{ck} + \frac{2\beta}{3\delta_{GB}^2}.
\end{align}

\begin{figure}[ht] 
	\centering
	\begin{subfigure}[b]{0.9\textwidth}
		\includegraphics[width=\textwidth]{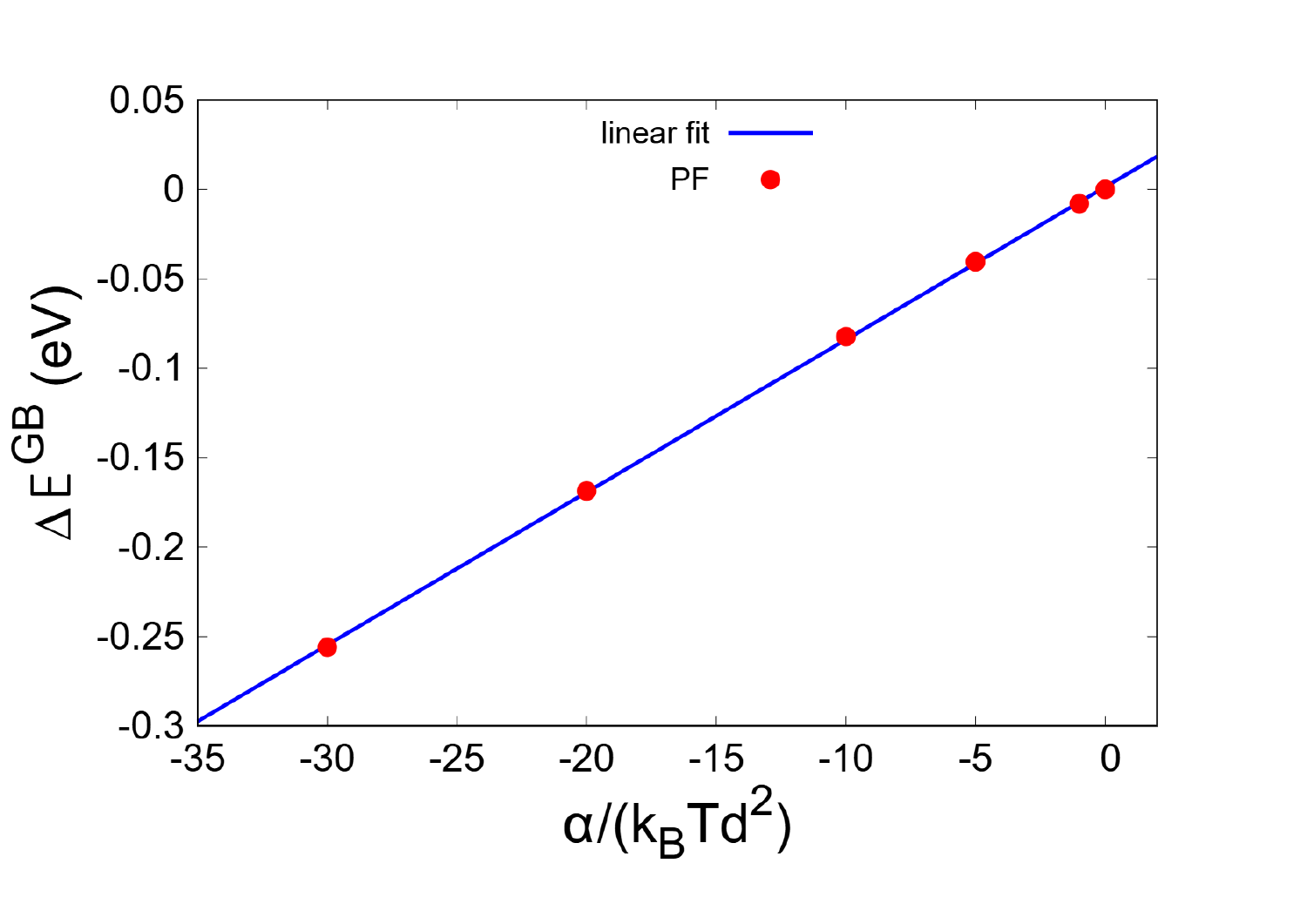}
		\caption{}
	\end{subfigure}%
	
	\begin{subfigure}[b]{0.9\textwidth}
		\includegraphics[width=\textwidth]{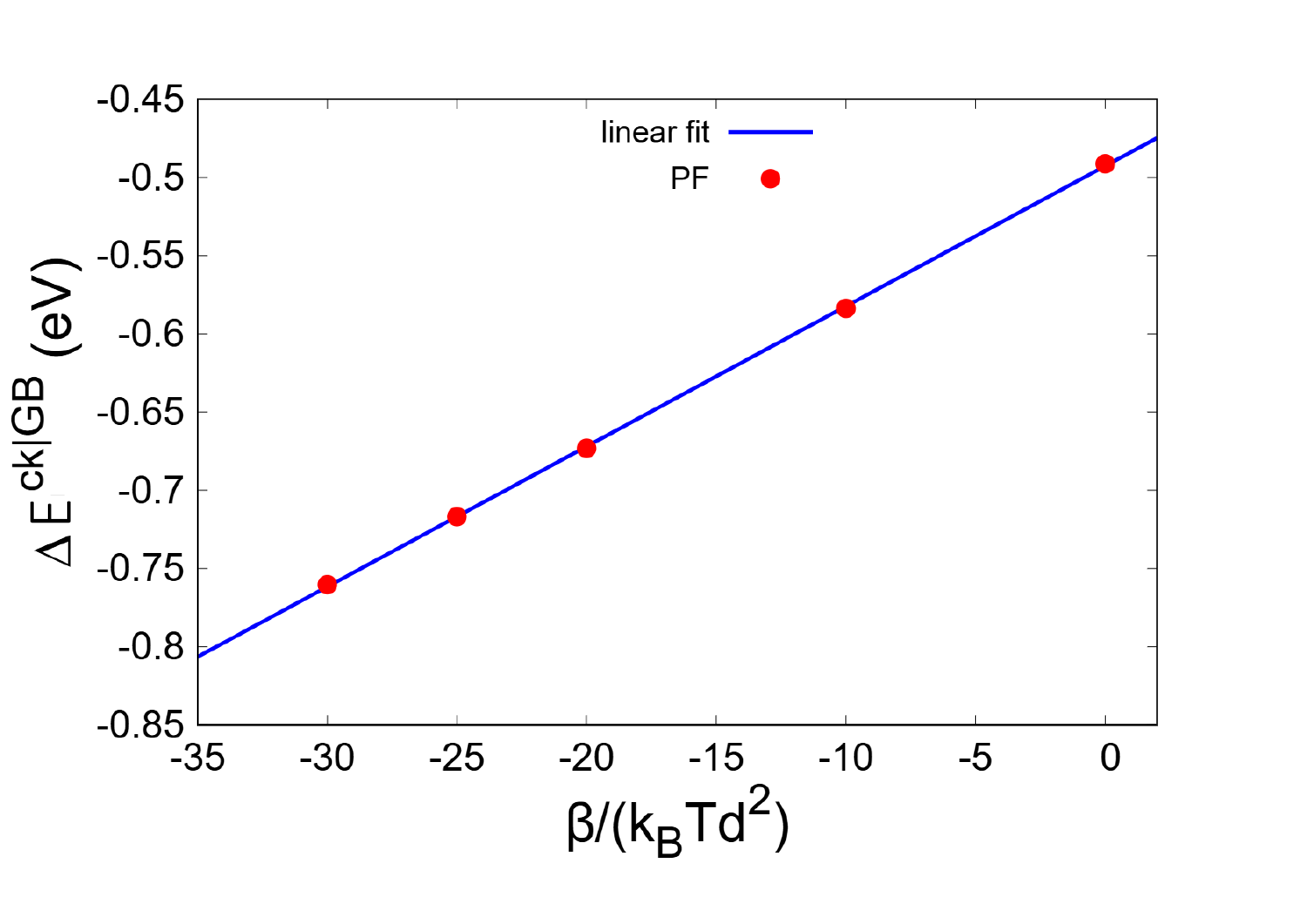}
		\caption{}
	\end{subfigure}
	\caption{Linear behavior of the hydrogen segregation energies in an intact and in a broken GB as a function of the parameters $\alpha$ and $\beta$.}
	\label{fig_alpha_beta}
\end{figure}

\noindent The hydrogen segregation energies in intact and broken GB, $\Delta E^{GB}$ and $\Delta E^{ck|GB}$ respectively, are found to be linear with $\alpha$ and $\beta$. When considering a spatially discretized simulation, the slopes of these linear relationships can be slightly different. 
In the present study, the slopes have been evaluated numerically by simulating the segregation mechanism in an intact and in a broken planar GB for different values of $\alpha$ and $\beta$. Using the PF parameter of our study ($\delta_{GB} = 2 d$ and $\Delta E^{ck} =$ -0.5 eV) we have obtained the results displayed in Fig.~\ref{fig_alpha_beta}.a and b, which confirm the linear behavior but with the slope that is 20\% smaller than the continuous solutions.  In our study, we used these numerical estimations by taking $\alpha/(k_{\scriptscriptstyle B} T d^2) = -30$ and $\beta/(k_{\scriptscriptstyle B} T d^2) = -45$ to fix $\Delta E^{GB} =$ -0.25 eV and $\Delta E^{ck|GB} =$ -0.9 eV, respectively.

\newpage

\section*{SIV: Analytical derivation of the hydrogen-induced surface energy variation $\Delta \gamma_{ch}$}

By definition, the effective surface energy in the presence of hydrogen is
\begin{equation}
	\gamma_{\mathrm{eff}}^{0} = \gamma^{0} + \Delta \gamma_{ch},
\end{equation}
where $\Delta \gamma_{ch}$ is the hydrogen-induced chemical surface energy variation. This quantity corresponds to the difference between chemical energies of the completely broken and intact states of the material, normalized by the crack surfaces area.

%$\mathcal{A}$:
%\begin{equation}
%	\Delta \gamma_{ch} = \frac{1}{2 \mathcal{A}} \left( \int_{V} f_{ch}  \mathrm{d} V - \int_{V} f_{ch}^{b,0}  \mathrm{d} V \right) 
%\end{equation}
%where $f_{ch}^{b,0}$ is the bulk chemical energy density, free of any crystal defects and $f_{ch}$ that of the system containing a crack.
%\begin{figure}[!h]
%	\centering
%	\begin{subfigure}{0.68\textwidth}
%		\includegraphics[width=\textwidth]{Sup_Mat_bulk_ck_system}
%	\end{subfigure} 
%	\caption*{}
%\end{figure}

This quantity can be computed for a planar crack, fully open so that no elastic energy remains in the system. By assuming that the crack volume fraction is negligible compared to the volume fraction of the bulk, we get
\begin{align}
	\Delta \gamma_{ch} & = \frac{1}{2} \left( \int_{-\infty}^{+\infty} h(\eta)f_{ch}^b(c^b) + (1-h(\eta))f_{ch}^{ck}(c^{ck})  \mathrm{d} V - \int_{-\infty}^{+\infty} f_{ch}^{b}  \mathrm{d} V \right).
\end{align}
In this expression, all fields are the equilibrium fields, but the subscript $eq$ is dropped for clarity. Within the KKS formalism used in this work, the homogeneity of the chemical potential $\mu = \frac{df}{dc}= \frac{df^b}{dc^b}= \frac{df^{ck}}{dc^{ck}}$ expected at equilibrium implies that, in this 1D geometry,  $c^b$ and $c^{ck}$ are also homogeneous quantities. Thus we get:
\begin{equation}
	\Delta \gamma_{ch}  = \left( f_{ch}^{ck}(c^{ck}) - f_{ch}^{b}(c^b) \right) \int_{0}^{+\infty} (1-h(\eta)) \mathrm d x.
	 \label{delta_gamma_ch_interm}
\end{equation}
In addition, it can be shown that $c^b$ and $c^{ck}$ correspond to the equilibrium values $c^b_{eq}$ and $c^{ck}_{eq}$ given by the common tangent construction,  and that the equilibrium $\eta$ profile  fulfills the relation:
\begin{equation}
 w(\eta) - \xi^2 \left(\frac{\mathrm d \eta}{\mathrm d x}\right)^2 = 0.
\end{equation}  
This allows us to write
\begin{equation}
\frac{\mathrm d \eta}{\sqrt{w(\eta)}} = \frac{\mathrm d x}{\xi}, 
\end{equation}
that can be used to perform a change of variable in Eq.~\eqref{delta_gamma_ch_interm}. We finally have
\begin{align}
	\Delta \gamma_{ch} &  = C_w^h \, \xi \, \left( f_{ch}^{ck,eq} - f_{ch}^{b,eq} \right), \label{delta_gamma_ch_exact}
\end{align}
where $f_{ch}^{ck,eq}= f_{ch}^{ck}(c^{ck}_{eq})$, $f_{ch}^{b,eq}= f_{ch}^{b}(c^{b}_{eq})$ and 
\begin{equation}
	C_w^h = \int_{0}^{1} \frac{1-h(\eta)}{\sqrt{w(\eta)}} \, d\eta,
\end{equation}
is an integration constant that only depends on functions $h$ and $w$. In our study, we used $h(\eta) = 3\eta^{2} - 2\eta^{3}$ and the KKL cohesion function $w(\eta) = 1 - 4\eta^{3} + 3\eta^{4}$ leading to $C_w^h = 0.5749$. 

\noindent Note finally that, in the limit of low concentrations, one has
\begin{align}
	f_{ch}^{\psi}(c^{\psi}) = \frac{c^{\psi}}{V_H} \left[E^{\psi} - k_{\scriptscriptstyle B} T \left( 1 -\ln c^{\psi} \right) \right].
\end{align}
By using these expressions in Eq. \eqref{delta_gamma_ch_exact} one finally gets
\begin{equation}
%	\Delta \gamma_{ch} = - k_B T C_w^h \frac{\xi}{V} \left( 1 - \exp \left[ \frac{\Delta E^{ck}}{k_B T} \right] \right) c^{ck,eq}
	\Delta \gamma_{ch} =  C_w^h \  f_{ch}^{b}  \ {\xi} \ \left( \exp \left[ -\frac{\Delta E^{ck}}{k_B T} \right] -1 \right). 
\end{equation}
In the case of hydrogen segregation at the surface considered in this work, this quantity is negative because both $f_{ch}^{b}$ and $\Delta E^{ck}$ are negative quantities.

\newpage

\section*{SV: Effective surface energy as a function of the hydrogen concentration at the surface}

\begin{figure}[ht]
	\centering
	\begin{subfigure}[b]{0.9\textwidth}
		\includegraphics[width=\textwidth]{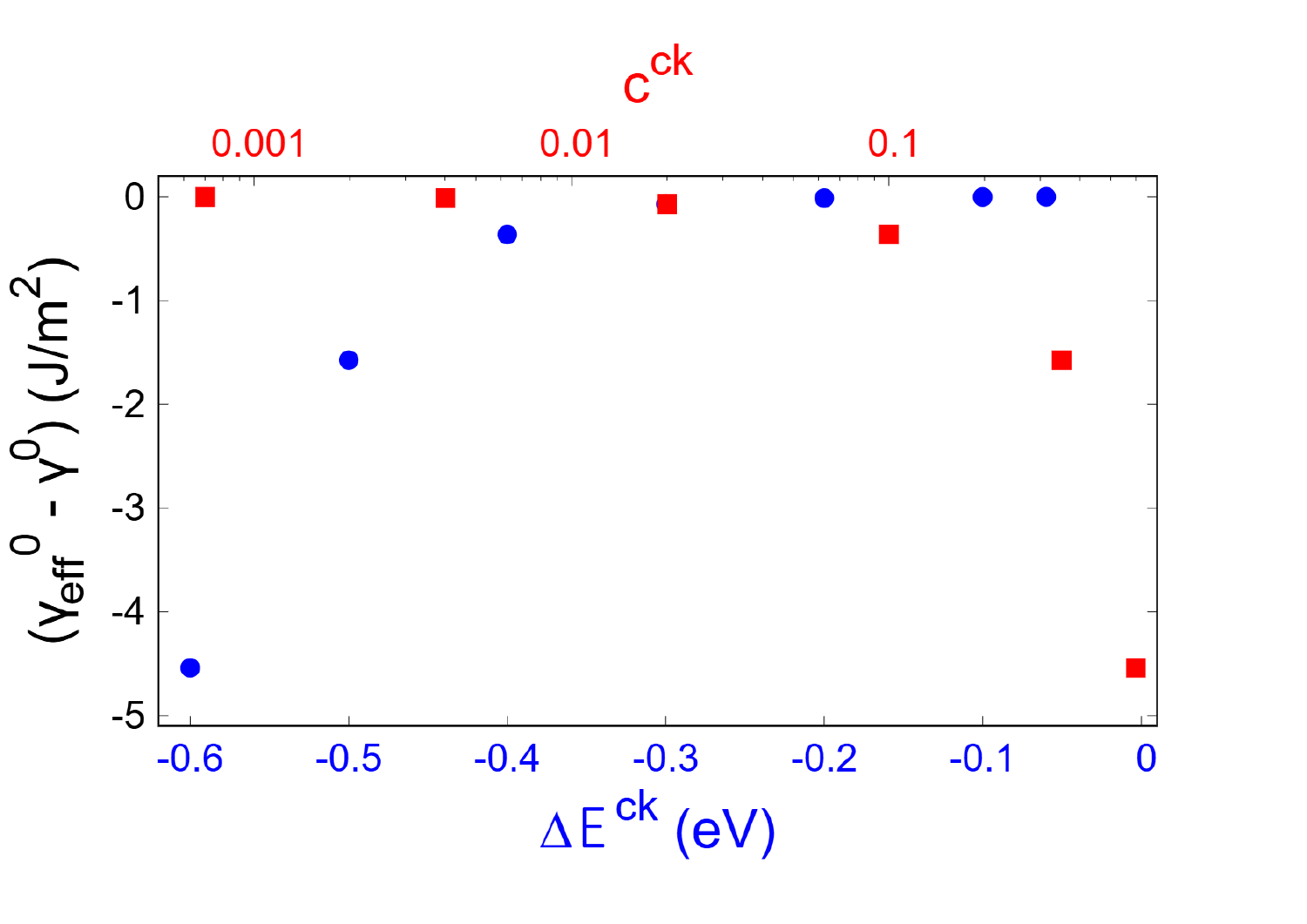}
		\caption{}
	\end{subfigure}
	
	\begin{subfigure}[b]{0.9\textwidth}
		\includegraphics[width=\textwidth]{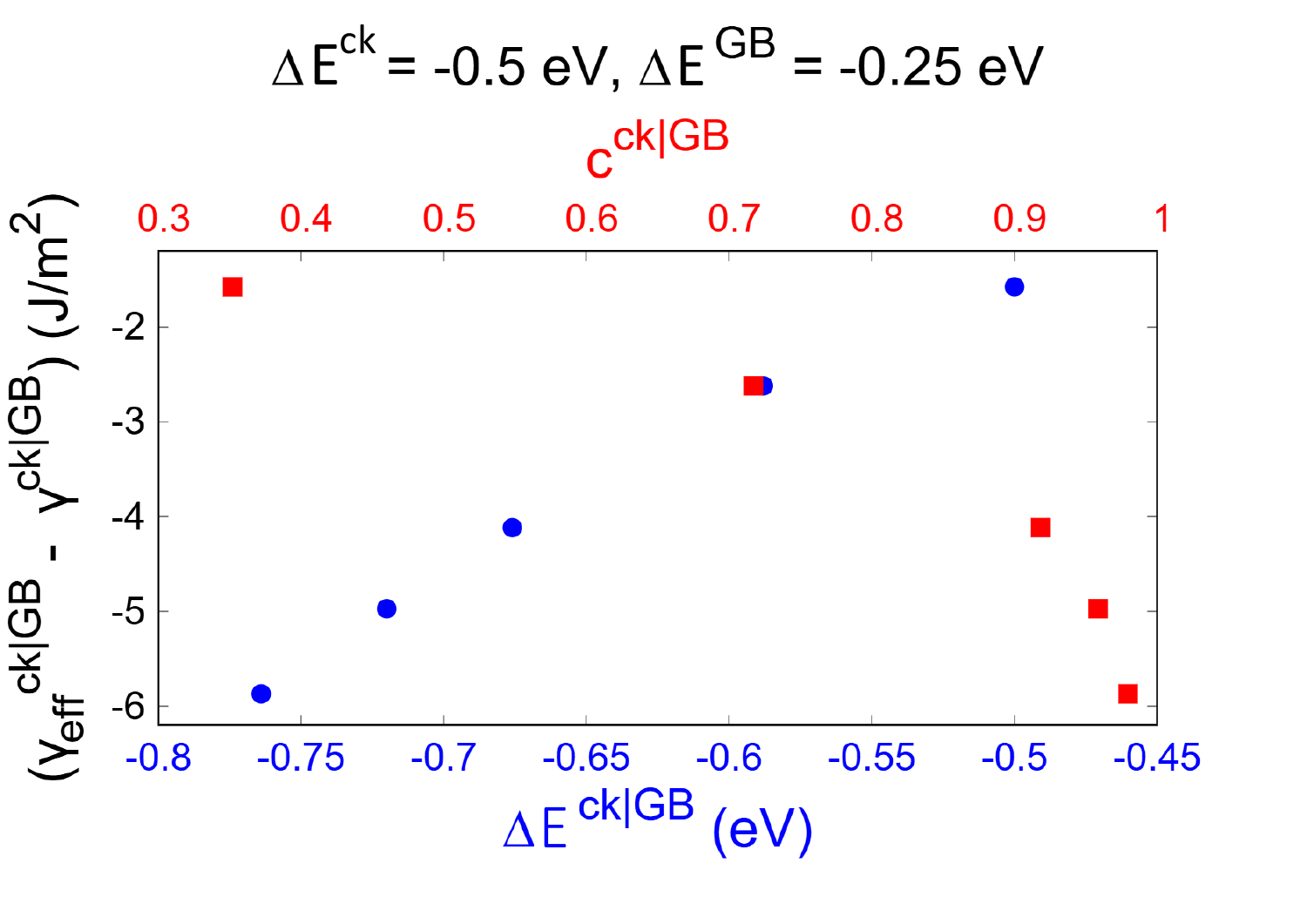}
		\caption{}
	\end{subfigure}
	\caption{Dependence of the effective surface energies with the hydrogen segregation energy or equivalently with the hydrogen coverage obtained from the PF simulations.}
	\label{fig_S6}
\end{figure}

As expected, both the hydrogen-induced surface energy variation $\Delta \gamma_{ch}$ and the hydrogen concentration at the surface $c^{ck}$ depend on the choice of the segregation energy $\Delta E^{ck}$.
These evolutions are shown in Fig.~\ref{fig_S6}-a.

\noindent Similarly, the hydrogen-induced surface energy variation at a cracked grain boundary is presented in  Fig.~\ref{fig_S6}-b together with the hydrogen concentration at the surface.
These quantities are drawn as a function of the effective segregation energy at the crack surface, for $\Delta E^{ck}=-0.5$eV and $\Delta E^{GB}=-0.25$eV.

%\newpage	
	
%\bibliographystyle{unsrturl}
%\bibliography{biblio_Sup_Mat}